\documentclass[prb,onecolumn,showpacs,preprintnumbers,amsmath,amssymb,floatfix]{revtex4}

\usepackage{graphicx}

\begin{document}

\title{Jastrow correlation factor for atoms, molecules, and solids}

\author{N.~D.~Drummond, M.~D.~Towler, and R.~J.~Needs}

\affiliation{TCM Group, Cavendish Laboratory, University of Cambridge,
Madingley Road, Cambridge CB3 0HE, UK}

\date{\today}

\begin{abstract}
A form of Jastrow factor is introduced for use in quantum Monte
Carlo simulations of finite and periodic systems.  Test data are
presented for atoms, molecules, and solids, including both
all-electron and pseudopotential atoms.  We demonstrate that our
Jastrow factor is able to retrieve a large fraction of the correlation
energy.
\end{abstract}


\maketitle

\section{Introduction}
\label{section:introduction}

Many-electron wave functions may be accurately and compactly
approximated by a product of a small number of Slater determinants and
a positive Jastrow correlation factor.  The Jastrow factor is an
explicit function of the electron--electron separations, so that
expectation values calculated with a Slater--Jastrow wave function do
not separate in the electron coordinates.  Nevertheless, the
variational and diffusion quantum Monte Carlo (VMC and DMC) methods
permit the use of such explicitly correlated wave functions.

In VMC, expectation values are calculated using an approximate trial
wave function, the integrals being performed by a Monte Carlo method.
In DMC~\cite{ceperley_1980,foulkes_2001} the imaginary-time
Schr\"odinger equation is used to evolve an ensemble of electronic
configurations towards the ground state.  The fermionic symmetry is
maintained by the fixed-node approximation,\cite{anderson_1976} in
which the nodal surface of the wave function is constrained to equal
that of a trial wave function.  The DMC method gives the energy that
would be obtained in a VMC calculation with the same Slater
determinants, but using the best possible Jastrow factor.

Although the DMC energy is in principle independent of the Jastrow
factor, a poor trial wave function increases the statistical error
bars and the time-step and population-control biases.  When non-local
pseudopotentials are used within DMC, the locality
approximation~\cite{hurley_1987,hammond_1987} leads to additional
errors which are second order in the error in the trial wave
function.\cite{mitas_1991}  The expectation values of operators that
do not commute with the Hamiltonian are often evaluated using
extrapolated estimation,\cite{foulkes_2001} the accuracy of the
extrapolation depending on the quality of the trial wave function.  In
practice the efficiency and accuracy of both VMC and DMC calculations
are critically dependent on the availability of high-quality Jastrow
factors.

Our Jastrow factor is designed for use in atoms, molecules, and
solids.  We have used it in a variety of systems, and here we report
results on the He, Ne$^{8+}$, Ne, and Ni atoms; the NiO and SiH$_4$
molecules; and crystalline Si in the diamond structure.  These systems
include all-electron and pseudopotential descriptions of atoms, with
the total number of electrons varying from 2 to 216.  We pay
particular attention to the issue of cutting off terms in the Jastrow
factor at finite ranges, which is desirable because of the local
nature of the inhomogeneous correlations in many systems, as well as
for reasons of computational efficiency in large systems.

We obtained the values of the free parameters in our Jastrow factors
by minimizing the variance of the
energy.\cite{umrigar_1988a,kent_1999}  All of our QMC calculations
were performed using the \textsc{casino} package.\cite{casino} We use
Hartree atomic units, $\hbar=|e|=m_e=4\pi \epsilon_0=1$, throughout
this article.

The rest of this paper is organized as follows.  In
Sec.~\ref{section:general_form} we describe the general form of our
Jastrow factor, while in Sec.~\ref{section:cusp_conditions} we show
how the electron--electron and electron--nucleus cusp
conditions~\cite{kato_pack} apply to this form.  The behavior of the
local energy at electron--electron and electron--nucleus coalescence
points is discussed in Sec.~\ref{section:E_L_at_coalescence_pts}.
Sec.~\ref{section:new_jastrow} describes the Jastrow factor in
detail.  In Sec.~\ref{section:further_comments} we make further
comments on the form of our Jastrow factor, while in
Sec.~\ref{section:specify_accuracy} we define our notation for
specifying the Jastrow factor and give our criterion for judging its
quality.  In Secs.~\ref{section:He_Ne}--\ref{section:Si} we report the
results of studies of various systems. Finally, we draw our
conclusions in Sec.~\ref{section:conclusions}.

\section{General form of the Jastrow factor}
\label{section:general_form}

The Slater--Jastrow wave function can be written as
\begin{equation}
\Psi(\{{\bf r}_i\},\{{\bf r}_I\}) = \exp\left[ J(\{{\bf r}_i\},\{{\bf
r}_I\}) \right] D(\{{\bf r}_i\})\;,
\end{equation}
where $\{{\bf r}_i\}$ and $\{{\bf r}_I\}$ denote the electron and ion
coordinates, respectively, $\exp\left[J\right]$ is the Jastrow factor,
and $D$ denotes the Slater part, which depends only implicitly on the
$\{{\bf r}_I\}$.

An accurate and efficient Jastrow factor should possess a number of
qualities.  The functional form of the Jastrow factor should be chosen
to reflect the physics of the correlations in the system, and it
should be parameterized efficiently.  The trial wave function must be
continuous everywhere and its gradient must be continuous wherever the
potential is finite, so that the kinetic energy is well-defined.  The
Kato cusp conditions~\cite{kato_pack} determine the behavior of the
many-body wave function when two electrons, or an electron and a
nucleus, are coincident.  The cusp conditions derive from the
requirement that the divergence in the local kinetic energy at a
coalescence point cancels the divergence in the local potential
energy.  Failure to satisfy the cusp conditions leads to divergences
in the local energy $\Psi^{-1} \hat{H} \Psi$, where $\hat{H}$ is the
Hamiltonian.  These divergences are especially harmful in DMC
calculations, where they can lead to population-control problems and
significant biases.  It is standard practice to use the Jastrow factor
to enforce the cusp conditions.  The Slater part of the wave function
is chosen to satisfy the correct symmetry under exchange of electrons,
and therefore the Jastrow factor should be symmetric under exchange.
Indeed the Slater part of the wave function is normally chosen to have
the correct symmetries of the state, so we should choose a Jastrow
factor that does not change this symmetry.  Finally, the Jastrow
factor should allow rapid evaluation, as this is one of the more
computationally demanding parts of VMC and DMC calculations.

Our Jastrow factor is the sum of homogeneous, isotropic
electron--electron terms $u$, isotropic electron--nucleus terms $\chi$
centered on the nuclei, isotropic electron--electron--nucleus terms
$f$, also centered on the nuclei and, in periodic systems, plane-wave
expansions of electron--electron separation and electron position, $p$
and $q$.  The form is
\begin{equation}
\label{eq:basic_J}
J(\{{\bf r}_i\},\{{\bf r}_I\}) = \sum_{i=1}^{N-1} \sum_{j=i+1}^N
u(r_{ij}) + \sum_{I=1}^{N_{\rm ions}} \sum_{i=1}^N \chi_I(r_{iI}) +
\sum_{I=1}^{N_{\rm ions}} \sum_{i=1}^{N-1} \sum_{j=i+1}^N
f_I(r_{iI},r_{jI},r_{ij}) +\sum_{i=1}^{N-1} \sum_{j=i+1}^N p({\bf
r}_{ij}) + \sum_{i=1}^N q({\bf r}_{i}),
\end{equation}
where $N$ is the number of electrons, $N_{\rm ions}$ is the number of
ions, ${\bf r}_{ij} = {\bf r}_{i} - {\bf r}_{j}$, and ${\bf r}_{iI} =
{\bf r}_{i} - {\bf r}_{I}$.  Note that $u$, $\chi$, $f$, $p$, and $q$
may also depend on the spins of $i$ and $j$.  Although we will present
results using spin-dependent parameters, for compactness the spin-type
has been suppressed in all formulae.  The basic form is not novel, as
terms of each type present in Eq.~(\ref{eq:basic_J}) have appeared in
Jastrow factors in the literature,\cite{foulkes_2001} but our
particular forms of $u$, $\chi_I$, and $f_I$ are new.

The plane-wave term, $p$, will describe similar sorts of correlation
to the $u$ term.  In periodic systems the $u$ term must be cut off at
a distance less than or equal to the Wigner--Seitz radius of the
simulation cell (see Sec.~\ref{section:jastrow_cutoffs}) and therefore
the $u$ function includes electron pairs over less than three quarters
of the simulation cell.  The $p$ term adds variational freedom in the
``corners'' of the simulation cell, which could be important in small
cells.  The $p$ term can also describe anisotropic correlations, such
as might be encountered in a layered compound.  However, we expect
that the $u$ term will be considerably more important than the $p$
term, which cannot describe the electron--electron cusps and is
therefore best limited to describing longer-ranged correlations.  The
$q$ term will describe similar electron--nucleus correlations to the
$\chi_I$ terms.

\section{The electron--electron and electron--nucleus cusp conditions}
\label{section:cusp_conditions}

Imposing the cusp conditions on the Jastrow factor is non-trivial
because the variables $r_{ij}$, $r_{iI}$, and $r_{jI}$ are not
independent.  It is important to understand the meaning of the
derivatives considered in this section.  In
Eq.~(\ref{eqn:par_kato_cusp}), for example, the derivative $\partial
\hat{\Psi} /\partial r_{ij}$ means the derivative with respect to
$r_{ij}$ with all other coordinates held fixed, while in
Eq.~(\ref{eqn:delta_Psi}) the derivative $\partial J /\partial r_{ij}$
means the derivative with respect to $r_{ij}$ with $r_i$ and $r_j$
fixed.

\subsection{The antiparallel-spin electron--electron cusp condition}

Consider the situation where two electrons of opposite spin, $i$ and
$j$, approach one another and the wave function is non-zero at the
two-particle coalescence point.  This condition holds at almost all
coalescence points of antiparallel-spin electrons.  Let us omit the
coordinates of all the other electrons and write the wave function in
terms of the center-of-mass and difference coordinates of electrons
$i$ and $j$, $\bar{\bf r}_{ij}=({\bf r}_i+{\bf r}_j)/2$ and ${\bf
r}_{ij}={\bf r}_i-{\bf r}_j$.  The cusp condition~\cite{kato_pack} is
\begin{equation}
\left( \frac{\partial \hat{\Psi}}{\partial r_{ij}} \right)_{r_{ij}=0}
= \frac{1}{2} \hat{\Psi}_{r_{ij}=0},
\label{eqn:par_kato_cusp}
\end{equation}
where $\hat{\Psi}(\bar{{\bf r}}_{ij},r_{ij})$ is the spherical average
of $\Psi(\bar{\bf r}_{ij},{\bf r}_{ij})$ about the coalescence point.

Neglecting the cuspless $p$ and $q$ terms, the Slater--Jastrow wave
function may be written as
\begin{equation}
\Psi(\bar{\bf r}_{ij},{\bf r}_{ij}) = \exp[J(r_{i},r_{j},r_{ij})]
  D(\bar{\bf r}_{ij},{\bf r}_{ij}),
\end{equation}
where for clarity we have assumed there is only one nucleus, which is
located at the origin. Consider the change in the value of $\Psi$ for
a small displacement from the coalescence point such that the center
of mass remains fixed:
\begin{equation}
\delta \Psi = \Psi_{r_{ij}=0} \times \left( \left[ \left(
  \frac{\partial J}{\partial r_{i}} \right) - \left( \frac{\partial
  J}{\partial r_{j}} \right) \right]_{r_{ij}=0} \delta r_{i} + \left(
  \frac{\partial J}{\partial r_{ij}} \right)_{r_{ij}=0} r_{ij} \right)
  + \exp[J_{r_{ij}=0}] \left( \nabla_{ij} D \right)_{r_{ij}=0} \cdot
  {\bf r}_{ij} + {\cal O}(r_{ij}^2),
\label{eqn:delta_Psi}
\end{equation}
where $\delta {\bf r}_i$ and $\delta {\bf r}_j$ are the changes in
${\bf r}_i$ and ${\bf r}_j$ when the electron separation ${\bf
r}_{ij}$ is increased from zero, and we have used $\delta{\bf
r}_j=-\delta {\bf r}_i$.  If the spherical average about the
coalescence point is taken then the terms involving $\delta r_{i}$ and
${\bf r}_{ij}$ vanish to ${\cal O}(r_{ij})$, so that
\begin{equation}
\delta \hat{\Psi} = \hat{\Psi}_{r_{ij}=0} \left( \frac{\partial
J}{\partial r_{ij}} \right)_{r_{ij}=0} r_{ij} + {\cal O}(r_{ij}^2).
\end{equation}
Hence the antiparallel cusp condition is equivalent to the requirement
that
\begin{equation}
\left( \frac{\partial J}{\partial r_{ij}} \right)_{r_{ij}=0 \atop
r_i=r_j}=\frac{1}{2},
\label{eqn:antipar_ee_jastrow_cusp}
\end{equation}
\noindent where $r_{ij}$, $r_i$, and $r_j$ are treated as independent
variables.

\subsection{The parallel-spin electron--electron cusp condition}

Suppose now that the approaching electrons $i$ and $j$ have parallel
spins.  The cusp condition~\cite{kato_pack} is
\begin{equation}
\left( \frac{\partial \Psi_{1m}}{\partial r_{ij}} \right)_{r_{ij}=0} =
\frac{1}{4} \left( \Psi_{1m} \right)_{r_{ij}=0},
\label{eqn:cusp_cond_wf_par}
\end{equation}
\noindent where $\Psi_{1m}$ is the $r_{ij} Y_{1m}$ component of
$\Psi$, and $Y_{lm}$ is the $(l,m)$th spherical harmonic.

Let us expand $\Psi$ about ${\bf r}_{ij}=0$.  $D$ is an odd function
of ${\bf r}_{ij}$; hence we obtain
\begin{equation}
\Psi = \exp[J_{r_{ij}=0}] \left( 1 + \left( \frac{\partial J}{\partial
r_{ij}} \right)_{r_{ij}=0} r_{ij} + \left[ \left( \frac{\partial
J}{\partial r_i} \right) - \left( \frac{\partial J}{\partial r_j}
\right) \right]_{r_{ij}=0} \delta r_i + {\cal O}(r_{ij}^2) \right)
\times \left( \left( \nabla_{ij} D \right)_{r_{ij}=0} \cdot {\bf
r}_{ij} + {\cal O}(r_{ij}^3) \right).
\end{equation}
The change in the electron--nucleus distance when the electron
separation $r_{ij}$ is increased from zero is $\delta r_i=r_{ij}
\cos(\theta_i)/2+{\cal O}(r_{ij}^2)$, where $\theta_i$ is the angle
between ${\bf r}_{ij}$ and ${\bf r}_i$.  The $r_{ij}Y_{1m}$ component
of $\Psi$ is therefore
\begin{equation}
\Psi_{1m} = \exp[J_{r_{ij}=0}] \left[ \left( \nabla_{ij} D
\right)_{r_{ij}=0} \cdot {\bf r}_{ij} \right]_{1m} \times \left( 1 +
\left( \frac{\partial J}{\partial r_{ij}} \right)_{r_{ij}=0} r_{ij}
+{\cal O}(r_{ij}^2) \right),
\end{equation}
where $\left[ X \right]_{1m}$ denotes the $r_{ij}Y_{1m}$ component of
$X$.  So the parallel-spin cusp condition of
Eq.~(\ref{eqn:cusp_cond_wf_par}) is equivalent to the requirement that
\begin{equation}
\left( \frac{\partial J}{\partial r_{ij}} \right)_{r_{ij}=0 \atop
r_i=r_j} = \frac{1}{4}, \label{eqn:par_ee_jastrow_cusp}
\end{equation}
where $r_{ij}$, $r_i$, and $r_j$ are treated as independent variables.

\subsection{The electron--nucleus cusp condition}

Now consider the cusp condition that must be satisfied as electron $i$
approaches a nucleus of atomic number $Z$.  The coordinates of all
other electrons are omitted.  The spherical average of $\Psi({\bf
r}_i)$ about the nucleus, $\bar{\Psi}(r_i)$, must obey~\cite{kato_pack}
\begin{equation}
\left( \frac{\partial \bar{\Psi}}{\partial r_{i}} \right)_{r_{i}=0} =
-Z \bar{\Psi}_{r_i=0}.
\end{equation}
By similar arguments to those given for the antiparallel
electron--electron cusp condition, if the Slater determinant is
continuously differentiable at the nucleus then the Jastrow factor
must satisfy
\begin{equation}
\left( \frac{\partial J}{\partial r_{i}} \right)_{r_{i}=0 \atop
r_{ij}=r_j}=-Z.
\label{eqn:nucleus_cusp}
\end{equation}
Note that if the Slater part of the wave function satisfies the
electron--nucleus cusp condition, or if a non-divergent
pseudopotential is used, then the Jastrow factor is required to be
cuspless at the nuclei: it should satisfy Eq.~(\ref{eqn:nucleus_cusp})
with $Z=0$.

\section{The behavior of the local energy at coalescence points
  \label{section:E_L_at_coalescence_pts}}

\subsection{Continuity at antiparallel-spin coalescence points}

The Slater--Jastrow wave function in the vicinity of an
antiparallel-spin coalescence point can be written as
\begin{equation}
\Psi({\bf r}_{ij}) = \exp[u(r_{ij})] S({\bf r}_{ij}), \label{eqn:wf_FS}
\end{equation}
where $S$ is the Slater wave function multiplied by the terms in the
Jastrow factor that are analytic at the coalescence point and we have
assumed there are no $f$ terms in the Jastrow factor.

Assuming that $u$ satisfies the Kato cusp condition of
Eq.~(\ref{eqn:antipar_ee_jastrow_cusp}), the local energy can be shown
to be
\begin{eqnarray}
E_L({\bf r}_{ij}) & = & -\frac{\nabla_{ij}^2 \Psi}{\Psi} +
  \frac{1}{r_{ij}} + E_{L0} \nonumber \\ & = & -\frac{1}{4} - 3 \left(
  \frac{d^2u}{dr_{ij}^2} \right)_{r_{ij}=0} - \frac{\nabla_{ij}^2
  S}{S} + E_{L0} - \frac{\left( \nabla_{ij} S \right)_{r_{ij}=0} \cdot
  {\bf r}_{ij}}{S_{r_{ij}=0} \times r_{ij}} + {\cal O}(r_{ij}),
  \label{eqn:El_at_antipar_CP}
\end{eqnarray}
where the $E_{L0}$ and $-S^{-1} \nabla_{ij}^2 S$ terms are continuous
  at the coalescence point.

\begin{figure}
\begin{center}
\includegraphics*{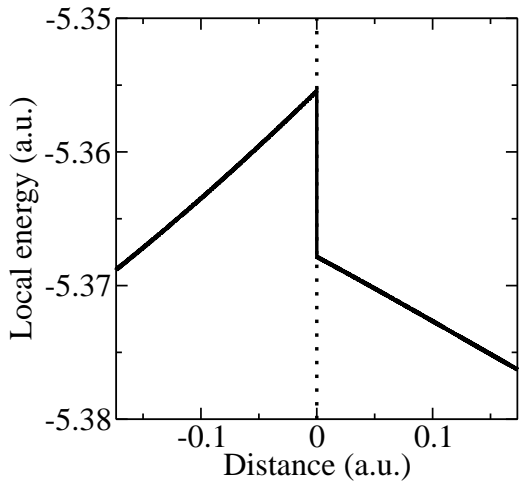}
\caption{Local energy plotted against the position of an electron as
it is moved along a straight line through another electron of the
opposite spin in SiH$_4$.  The dotted line indicates the location of
the stationary electron.
\label{fig:silane_EL_antiparallel}}
\end{center}
\end{figure}

Satisfying the cusp condition removes the divergence in the local
energy at the coalescence point, irrespective of the angle at which
the electrons approach.  However, the ${\cal O}(r_{ij}^0)$ term in
Eq.~(\ref{eqn:El_at_antipar_CP}) \textit{does} depend on the direction
of approach.  The local energy therefore has a point discontinuity at
antiparallel-spin coalescence points.  This behavior is illustrated in
Fig.~\ref{fig:silane_EL_antiparallel}.

There is a similar discontinuity in the local energy at nuclei when
the electron--nucleus cusp condition is enforced.  If, on the other
hand, the no-cusp condition is enforced at the center of a
pseudo-atom, there is no discontinuity in the local energy.

\subsection{Continuity at parallel-spin coalescence points}

\begin{figure}
\begin{center}
\includegraphics*{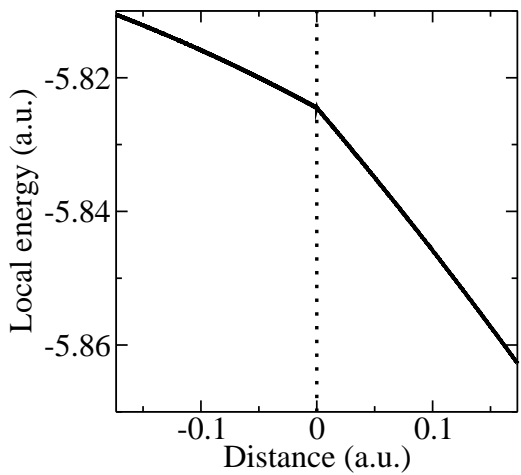}
\caption{Local energy plotted against the position of an electron as
it is moved along a straight line through another electron of the same
spin in SiH$_4$.  The dotted line indicates the location of the
stationary electron.
\label{fig:silane_EL_parallel}}
\end{center}
\end{figure}

Now consider a parallel-spin coalescence point.  Again the wave
function may be written in the form of Eq.~(\ref{eqn:wf_FS}) in the
vicinity of the coalescence point, but this time $S$ is an odd
function of ${\bf r}_{ij}$.  If the Kato cusp condition of
Eq.~(\ref{eqn:par_ee_jastrow_cusp}) is satisfied by $u$, the local
energy is
\begin{equation}
E_L = -\frac{1}{16} - 5\left( \frac{d^2u}{dr_{ij}^2}
    \right)_{r_{ij}=0} - \frac{\nabla_{ij}^2 S}{S} + E_{L0} + {\cal
    O}(r_{ij}).
\end{equation}
The $-S^{-1} \nabla_{ij}^2 S$ term is discontinuous at a parallel-spin
coalescence point, giving a point discontinuity in the local energy.
In spite of this, the local energy is continuous when one electron is
moved along a straight line through another of the same spin because
of the symmetry of the local energy with respect to exchanges of
parallel-spin electrons.  This behavior is illustrated in
Fig.~\ref{fig:silane_EL_parallel}.

\subsection{Further coalescence conditions}

Rassolov and Chipman \cite{rassolov} have demonstrated that, at the
coalescence point of two antiparallel-spin electrons $i$ and $j$,
\begin{equation}
\left( \frac{\partial^3 \hat{\Psi}}{\partial r_{ij}^3}
\right)_{r_{ij}=0} = \left( \frac{\partial^2 \hat{\Psi}}{\partial
r_{ij}^2} \right)_{r_{ij}=0} - \frac{\hat{\Psi}_{r_{ij}=0}}{8},
\label{eqn:rassolov_antiparallel}
\end{equation}
while for two parallel-spin electrons
\begin{equation}
\left( \frac{\partial^3 \Psi_{1m}}{\partial r_{ij}^3}
\right)_{r_{ij}=0} = \frac{7}{12} \left( \frac{\partial^2
\Psi_{1m}}{\partial r_{ij}^2} \right)_{r_{ij}=0} - \frac{\left(
\Psi_{1m} \right)_{r_{ij}=0}}{48}.
\label{eqn:rassolov_parallel}
\end{equation}

These cusp conditions are difficult to apply to Slater--Jastrow wave
functions because Eqs.~(\ref{eqn:rassolov_antiparallel}) and
(\ref{eqn:rassolov_parallel}) involve the Slater determinant as well
as the Jastrow factor.  If we assume that $\left( \partial^2 \hat{S} /
\partial r_{ij}^2 \right)_{r_{ij}=0}=0$,
\noindent where $\hat{S}$ is the spherical average of $S$ about an
antiparallel-spin coalescence point, then we can derive an approximate
condition on the antiparallel-spin $u$ term in the Jastrow
factor.\cite{ndd_thesis}  Likewise, if we assume that $\left(
\partial^2 S_{1m} / \partial r_{ij}^2 \right)_{r_{ij}=0}=0$, where
$S_{1m}$ is the $r_{ij} Y_{1m}$ component of $S$ about a parallel-spin
coalescence point, then we can derive an approximate condition on the
parallel-spin $u$ term.  Imposing these additional conditions was not
found to be of any benefit in practice.

The Rassolov--Chipman conditions can be derived by demanding that
$\hat{\Psi}^{-1} \hat{H} \hat{\Psi}$ and $\left( \Psi_{1m} r_{ij}
Y_{1m} \right)^{-1} \hat{H} \left( \Psi_{1m} r_{ij} Y_{1m} \right)$
are cuspless at antiparallel- and parallel-spin coalescence points,
respectively.\cite{ndd_thesis} There seems little point in attempting
to apply the Rassolov--Chipman conditions unless one has already
ensured that the local energy itself is continuous at coalescence
points.  A continuous local energy can be achieved in a two-electron
atom by using a trial wave function based upon the Fock
expansion.\cite{myers_1991} However, it is unlikely that a practical
method for eliminating the local-energy discontinuities in larger
systems will be found.

\section{The Jastrow factor \label{section:new_jastrow}}

\subsection{The $u$, $\chi$, and $f$ terms}

For the $u$ term we use an expression which is a variation on the form
we have used for a number of years~\cite{williamson_1996} and consists
of a complete power expansion in $r_{ij}$ up to order $r_{ij}^{C+N_u}$
which satisfies the Kato cusp conditions at $r_{ij}=0$, goes to zero
at the cutoff length, $r_{ij}=L_u$, and has $C-1$ continuous
derivatives at $L_u$:
\begin{equation}
u(r_{ij}) = (r_{ij}-L_u)^C \Theta(L_u-r_{ij}) \times \left( \alpha_0 +
\left[ \frac{\Gamma_{ij}}{(-L_u)^C} + \frac{\alpha_0 C}{L_u} \right]
r_{ij} + \sum_{l=2}^{N_u} \alpha_l r_{ij}^l \right),
\label{eqn:u_defn}
\end{equation}
\noindent where $\Theta$ is the Heaviside function and
$\Gamma_{ij}=1/2$ if electrons $i$ and $j$ have opposite spins and
$\Gamma_{ij}=1/4$ if $i$ and $j$ have the same spin.  In this
expression $C$ determines the behavior at the cutoff length.  If
$C=2$, the gradient of $u$ is continuous but the second derivative and
hence the local energy is discontinuous, and if $C=3$ then both the
gradient of $u$ and the local energy are continuous.

The form of $\chi$ is similarly related to our earlier work,
\begin{equation}
\chi_I(r_{iI}) = (r_{iI}-L_{\chi I})^C \Theta(L_{\chi I}-r_{iI}) {}
\times \left( \beta_{0I} + \left[\frac{-Z_I}{(-L_{\chi I})^C} +
\frac{\beta_{0I} C}{L_{\chi I}} \right] r_{iI} + \sum_{m=2}^{N_\chi}
\beta_{mI} r_{iI}^m \right). \label{eqn:chi_defn}
\end{equation}
It may be assumed that $\beta_{mI}=\beta_{mJ}$ where $I$ and $J$ are
equivalent ions.  The term involving the ionic charge $Z_I$ enforces
the electron--nucleus cusp condition.

The expression for $f$ is the most general expansion of a function of
$r_{ij}$, $r_{iI}$, and $r_{jI}$ that is cuspless at the coalescence
point and goes smoothly to zero when either $r_{iI}$ or $r_{jI}$ reach
cutoff lengths:
\begin{equation}
f_I(r_{iI},r_{jI},r_{ij}) = (r_{iI}-L_{f I})^C \Theta(L_{f I}-r_{iI})
\times (r_{jI}-L_{f I})^C \Theta(L_{f I}-r_{jI})  \times
\sum_{l=0}^{N_{f I}^{\rm eN}} \sum_{m=0}^{N_{f I}^{\rm eN}}
\sum_{n=0}^{N_{f I}^{\rm ee}} \gamma_{lmnI} r_{iI}^l r_{jI}^m r_{ij}^n.
\label{eqn:f_defn}
\end{equation}
Various restrictions are placed on $\gamma_{lmnI}$.  To ensure the
Jastrow factor is symmetric under electron exchanges we demand that
$\gamma_{lmnI} =\gamma_{mlnI}$ $\forall \, I,m,l,n$.  If ions $I$ and
$J$ are equivalent then we demand that $\gamma_{lmnI}=\gamma_{lmnJ}$.
The condition that the $f$ term has no electron--electron cusps is
\begin{equation}
\left( \frac{\partial f}{\partial r_{ij}} \right)_{r_{ij}=0 \atop
  r_{iI}=r_{jI}} = 0,
\end{equation}
which implies that
\begin{equation}
\sum_{l=0}^{N_{f I}^{\rm eN}} \sum_{m=0}^{N_{f I}^{\rm eN}}
      \gamma_{lm1I} r_{iI}^{l+m} (r_{iI}-L_{f I})^{2C} = 0\;,
\end{equation}
for all $r_{iI}$.  Hence, $\forall \, k \in \{0,\ldots,2N_{f I}^{\rm
eN} \}$, we must have
\begin{equation}
\sum_{l,m {\rm ~:~} l+m=k} \gamma_{lm1I}=0.
\label{equation:gamma_cond_ee}
\end{equation}
The condition that the $f$ term has no electron--nucleus cusps is
\begin{equation}
\left( \frac{\partial f}{\partial r_{iI}} \right)_{r_{iI}=0 \atop
r_{ij}=r_{jI}} = 0,
\end{equation}
which gives
\begin{equation}
\sum_{m=0}^{N_{f I}^{\rm eN}} \sum_{n=0}^{N_{f I}^{\rm ee}} (
C\gamma_{0mnI}-L_{f I}\gamma_{1mnI}) \times (-L_{f I})^{C-1}
r_{jI}^{m+n}(r_{jI}-L_{f I})^C = 0,
\end{equation}
for all $r_{jI}$.  We therefore require that, $\forall \, k^\prime \in
\{0,\ldots,N_{f I}^{\rm eN}+N_{f I}^{\rm ee} \}$,
\begin{equation}
\sum_{m,n {\rm ~:~} m+n=k^\prime} (C \gamma_{0mnI} - L_{f I}
\gamma_{1mnI})=0.
\label{equation:gamma_cond_eN}
\end{equation}

The method by which we impose the various constraints is described in
detail in Appendix~\ref{appendix:constraints}.

\subsection{The $p$ and $q$ terms}

The $p$ term takes the cuspless form
\begin{equation}
p({\bf r}_{ij}) = \sum_A a_A \sum_{{\bf G}_A^+} \cos({\bf G}_A \cdot
  {\bf r}_{ij})\;,
\end{equation}
where the $\{ {\bf G}_A \}$ are the reciprocal lattice vectors of the
simulation cell belonging to the $A$th star of vectors that are
equivalent under the full symmetry group of the Bravais lattice, and
``$+$'' means that, if ${\bf G}_A$ is included in the sum, $-{\bf
G}_A$ is excluded.

For systems with inversion symmetry the $q$ term takes the cuspless
form
\begin{equation}
q({\bf r}_{i}) = \sum_B b_{B} \sum_{{\bf G}_B^+} \cos({\bf G}_B \cdot
{\bf r}_{i}),
\end{equation}
where the $\{ {\bf G}_B \}$ are the reciprocal lattice vectors of the
primitive unit cell belonging to the $B$th star of vectors that are
equivalent under the space-group symmetry of the crystal, and the
``$+$'' means that, if ${\bf G}_B$ is included in the sum, $-{\bf
G}_B$ is excluded.  In this work $q$ has only been used for systems
with inversion symmetry.  Plane-wave expansions of electron position
can also be used for systems without inversion
symmetry.\cite{malatesta}

\subsection{Cutting off terms in the Jastrow factor
  \label{section:jastrow_cutoffs}}

To avoid unwanted derivative discontinuities in the wave function of a
periodic system, the isotropic functions $u$, $\chi_I$, and $f_I$ must
be cut off at a distance less than or equal to the Wigner--Seitz
radius of the simulation cell.  Furthermore, reasons of efficiency
dictate that in particular $f_I$ should be cut off at short distances
in both finite and periodic systems.  Suppose we wish to evaluate a
Slater--Jastrow wave function for a number of systems of increasing
size, where the number of electrons $N$ is assumed to be proportional
to the number of ions, $N_{\rm ions}$.  If the cutoff lengths $L_u$,
$L_{\chi}$, and $L_f$ are chosen to be proportional to the size of the
system then the numbers of operations required to update $u$ and
$\chi_I$ after each electron move are ${\cal O}(N)$.  The cost of
updating the $f_I$ term is, however, proportional to ${\cal O}(N^2)$,
which is prohibitive in large systems.  If we choose the cutoff
lengths to be independent of the system size then each term can be
updated in ${\cal O}(1)$ operations.

The $f_I$ term describes inhomogeneous correlations that are
spherically symmetric about atom $I$.  It does not seem likely that
$f_I$ could describe the inhomogeneity in correlations at points
distant from atom $I$ in systems with many atoms.  Similarly we argue
that the long-ranged part of the spherically symmetric $\chi_I$ terms
will not give useful variational freedom.  For a fixed number of
variable parameters we therefore expect that the best results will be
obtained by cutting off the $\chi_I$ and $f_I$ terms at distances of
roughly the size of atom $I$, so that the variational freedom in these
terms is concentrated at short distances where it is most useful.  The
$u$ term must describe both long- and short-ranged correlations and
therefore we expect it to be long-ranged.  In our implementation we
allow the cutoff lengths $L_u$, $L_{\chi}$, and $L_f$ to be varied,
and these degrees of freedom are investigated in
Secs.~\ref{section:He_Ne}--\ref{section:Si}.  In many cases the
optimal value of $L_u$ is approximately proportional to the system
size and the optimal values of $L_{\chi}$ and $L_f$ are approximately
independent of the system size, so that overall the cost of updating
the Jastrow factor after each electron move scales as ${\cal O}(N)$.

As mentioned earlier, the value of $C$ in
Eqs.~(\ref{eqn:u_defn}),~(\ref{eqn:chi_defn}), and~(\ref{eqn:f_defn})
determines the behavior of the Jastrow factor, and hence the local
energy, at the cutoff lengths.  Discontinuities in the local energy at
the cutoff lengths may be harmful to optimization procedures, but the
price paid for having a smoother local energy is a reduction in
variational freedom.

\section{Further comments on our Jastrow factor}
\label{section:further_comments}

We have used power series in the inter-particle distances rather than
scaled variables, such as $r_{ij}/(1+br_{ij})$, which have been used
by Boys and Handy\cite{boys_3} and
others.\cite{schmidt_1990,filippi_1996} These scaled variables go to a
constant at large $r_{ij}$, which is useful in finite systems.
However, it is not clear whether scaled variables are helpful when the
$u$, $\chi$, and $f$ terms are cut off at finite lengths, as they must
be in periodic systems.

In our previous Jastrow factors~\cite{williamson_1996} we used
Chebyshev polynomials rather than the powers themselves.  The ideas
behind this were that (i) the Chebyshev series may be calculated to
very high accuracy with double-precision arithmetic using recurrence
relations and (ii) the optimal coefficients tend to be of a more
uniform magnitude, which could be helpful within optimization
procedures.  However, we have found that the precision offered by a
simple power series with double-precision arithmetic is perfectly
acceptable up to an order of at least 20, and we have found no clear
benefits from the use of Chebyshev polynomials within our current
optimization procedures either.  We have therefore chosen to use
simple power series, which may be evaluated more rapidly than the
corresponding Chebyshev series.

Our new Jastrow factor includes terms such as $r_i h(r_j,r_{ij})$, and
$r_{ij} g(r_i,r_j)$, where $h$ and $g$ are polynomials, which are
absent in the Jastrow factors used by Schmidt and
Moskowitz~\cite{schmidt_1990,moskowitz_1992} and some other
researchers. Such terms do not in general satisfy the cusp conditions
on their own, but certain linear combinations of them do and therefore
they should be allowed to occur in the power series. The Jastrow
factors used by Umrigar and
coworkers~\cite{umrigar_1988b,filippi_1996,huang_1997} have included
such terms for many years.  We report tests of the importance of the
terms neglected in the Schmidt--Moskowitz Jastrow factor in
Sec.~\ref{section:He_Ne}.

As an option within our implementation we may try to reduce the extent
to which $f$ duplicates the $u$ and $\chi$ terms in the Jastrow factor
by imposing the conditions $\gamma_{00nI}=0$ for all $n$ and
$\gamma_{l00I}=0$ for all $l$.  Note, however, that the terms of $f$
with $l=m=0$ do not exactly correspond to $u$: they are
electron--electron terms local to ion $I$.  This variational freedom,
which is investigated in Sec.~\ref{section:He_Ne}, may be used to
describe correlations that occur on two different length scales, for
example, in the core and valence electrons of an atom or the intra-
and inter-atomic electron correlations of weakly interacting atoms.
The terms of $f$ with $m=n=0$ are less likely to give useful
variational freedom.  However, the use of duplication of $u$ and
$\chi$ by $f$ does not appear to cause any difficulties within our
optimization procedure, even where $L_\chi \simeq L_f$.

The variable parameters appear linearly in our Jastrow factor, with
the exception of the cutoff lengths, $L_u$, $L_{\chi I}$, and $L_{f
I}$.  The least-squares function in an unreweighted variance
minimization is quartic in the linear parameters;\cite{kent_1999}
however the dependence on the cutoff lengths is much more complicated,
and they couple strongly to the other parameters.  The use of linear
parameters is found to be very advantageous in practice: up to ten
times fewer Gauss--Newton iterations are required to converge to the
minimum of the least-squares function when the cutoff lengths are
fixed compared with when they are optimized.

Our Jastrow factor does not include logarithmic terms such as those
motivated by the Fock expansion.\cite{fock_1958,morgan_1986} Although
these terms have been used in highly accurate Hylleraas-expansion
calculations for two-electron atoms~\cite{myers_1991} it should be
noted that the most accurate calculations of this type performed to
date have not included them.\cite{drake_2002} It cannot therefore be
necessary to include the logarithmic terms to obtain high accuracy.

Our Jastrow factor does not include terms involving three or more
electrons.  Of course the repulsive Coulomb interaction and the
antisymmetry of the wave function ensures that three or more electrons
rarely come close to one another, so that such terms are expected to
be small, and explicit tests by Huang \textit{et
al.}~\cite{huang_1997} support this view.

In our implementation it is possible to use different $u$, $f$, and
$p$ functions for antiparallel, parallel spin-up, and parallel
spin-down pairs of electrons, and different $\chi$ and $q$ functions
for spin-up and spin-down electrons.  We investigate the effect of
using the different possible spin-dependences in the Jastrow factor
for a partially polarized system in Sec.~\ref{section:Ni}.  Note that
if different $u$ functions are used for parallel- and
antiparallel-spin pairs of electrons then both of the Kato cusp
conditions are satisfied, but if the same $u$ function is used for all
pairs of electrons then only the antiparallel-spin cusp condition is
satisfied.  The use of a Jastrow factor that is not symmetric with
respect to exchanges of electrons of opposite spin generally produces
a trial wave function that is not an eigenfunction of the spin
operator $\hat{S}^2$, even though the ground-state wave function must
be an eigenfunction of $\hat{S}^2$.  An investigation into this
\textit{spin-contamination} effect has been carried out by Huang
\textit{et al.}~\cite{huang_1998}, who found that highly optimized
wave functions suffer from relatively little spin contamination.

\section{Specification of the Jastrow factor and the measure of accuracy}
\label{section:specify_accuracy}

In the tests reported here the parameter $C$, which determines the
behavior at the cutoff lengths, takes the values $C=2$ or $C=3$.  The
terms included in $u$ are specified by $N_u$, those in $\chi$ by
$N_\chi$, and those in $f$ by $N_f^{\rm eN}$ and $N_f^{\rm ee}$.  In
each case $N_\chi$ is the same for all species of atom present, and
likewise for $N_f^{\rm eN}$ and $N_f^{\rm ee}$.  Spin dependences in
$u$, $\chi$, $f$, $p$, and $q$ are specified by $S_u$, $S_{\chi}$,
$S_f$, $S_p$, and $S_q$, where $S_u=0$ denotes that the same $u$
function is used for parallel- and antiparallel-spin pairs, $S_u=1$
denotes that different functions are used for parallel- and
antiparallel-spin pairs, and $S_u=2$ denotes that different functions
are used for parallel spin-up, parallel spin-down and
antiparallel-spin pairs.  $S_{\chi}=0$ denotes that the same $\chi$
function is used for spin-up and spin-down electrons while $S_{\chi} =
1$ if they are allowed to be different, etc.  Duplication of the
$r_{ij}$ terms in $u$ and the $r_i$ terms in $\chi$ by $f$ is denoted
by ``D''~=~True.  The terms included in the plane-wave expansions, $p$
and $q$, are determined by the number of stars of ${\bf G}$ vectors
included, $N_p$ and $N_q$.  If ``SMJ''~=~True then only the $f$ terms
contained in the Schmidt--Moskowitz Jastrow factor are used (i.e., the
terms proportional to $r_{ij}$ and $r_i$ are omitted).  In each case
we will specify the relevant descriptors and give the total number of
optimized parameters in the Jastrow factor, $N_T$.  The cutoff lengths
are included in the count of parameters.

Unless otherwise stated, the $\chi_I$ functions were chosen to be
cuspless at the nuclei (i.e.\ $Z_I=0$ in each case) because
non-divergent pseudopotentials were used or, where the full Coulomb
potential was used, the orbitals satisfied the electron--nucleus cusp
condition.

To initiate the optimization procedure one must select a set of
configurations from a suitable probability distribution.  We have
found that the distribution obtained from the square of the Slater
part of the wave function is normally an excellent starting point;
indeed our results suggest that it may be preferable to the
``self-consistent'' approach of updating the distribution to include
the latest estimate of the Jastrow factor.  It should be noted that if
one sets all of the variable parameters in our Jastrow factor to zero,
the resulting wave function can be very poor, often giving energies
which are higher than that obtained using the Slater part only.

We measure the accuracy of a Jastrow factor by the percentage of the
DMC correlation energy retrieved within VMC, i.e.,
\begin{equation}
\eta = \frac{E_{\rm HF}-E_{\rm VMC}}{E_{\rm HF}-E_{\rm DMC}} \times
100 \% \;,
\end{equation}
where $E_{\rm HF}$ is the energy obtained with the Slater determinants
only, $E_{\rm VMC}$ is the VMC energy obtained with the
Slater--Jastrow wave function, and $E_{\rm DMC}$ is the DMC energy.
The DMC method gives the energy corresponding to a perfect Jastrow
factor: see Sec.~\ref{section:introduction}.  In this work the
orbitals in the Slater determinants were kept fixed and we only
optimized the Jastrow factor.  Under these conditions $\eta$ is an
appropriate measure of the accuracy of Jastrow factors. We also report
the variance of the local energy, $\sigma_E^2$, for each Jastrow
factor tested.  The energy variance is the quantity that determines
the size of the statistical error bars for a given computational
effort in QMC calculations.  Furthermore, it is the object that we
actually minimize when optimizing the Jastrow factor.

\section{Example I: He and Ne atoms}
\label{section:He_Ne}

\subsection{Two-electron atoms}

Extremely accurate energies are available for the two-electron He and
Ne$^{8+}$ atoms from variational calculations using Hylleraas
expansions~\cite{drake_2002} and other methods.  It is straightforward
to show that the exact ground-state wave function of a two-electron
atom is a nodeless function of $r_1$, $r_2$, and $r_{12}$.  It should
therefore be possible to obtain very accurate results by including
$f(r_1,r_2,r_{12})$ terms in the Jastrow factor. As the ground-state
wave function is nodeless, the DMC energy should equal the exact
(non-relativistic and infinite-nuclear-mass) energy, apart from
statistical errors and biases due to the use of finite time steps and
populations.  We used orbitals derived from numerical integrations of
the Hartree--Fock (HF) equations on fine radial grids.

\begin{table}
\begin{center}
\begin{tabular}{ccccccr@{.}lr@{.}lr@{.}l}
\hline \hline

$C$ & $N_{u,\chi}$ & $N_f^{\rm eN,ee}$ & D & SMJ & $N_T$ &
\multicolumn{2}{c}{$E_{\rm VMC}$ (a.u.)} & \multicolumn{2}{c}{$\eta$}
& \multicolumn{2}{c}{$\sigma_E^2$ (a.u.)} \\

\hline

3 & 8 & 0 & -- & -- & 18 & $-2$&$900010(9)$ & $91$&$17(2)$\%  &
$0$&$0237(4)$ \\

3 & 6 & 3 & T  & T  & 33 & $-2$&$903555(2)$ & $99$&$598(5)$\% &
$0$&$002450(6)$ \\

2 & 8 & 3 & F  & F  & 40 & $-2$&$903596(2)$ & $99$&$696(5)$\% &
$0$&$00246(6)$ \\

3 & 6 & 3 & F  & F  & 36 & $-2$&$903660(3)$ & $99$&$848(7)$\% &
$0$&$00083(1)$ \\

3 & 6 & 3 & T  & F  & 41 & $-2$&$903693(1)$ & $99$&$926(2)$\% &
$0$&$000653(4)$ \\

\hline\hline
\end{tabular}
\caption{Optimized Jastrow factors and VMC energies for He.  The HF
energy is $-2.86167999$\,a.u., the exact energy is $-2.903724$\,a.u.,
and the DMC energy is within error bars of the exact value.  In each
case $S_u=S_f=S_\chi=0$, $N_u=N_\chi \equiv N_{u,\chi}$, and $N_f^{\rm
eN}=N_f^{\rm ee} \equiv N_F^{\rm eN,ee}$.
\label{table:varmin_results_He}}
\end{center}
\end{table}

\begin{table}
\begin{center}
\begin{tabular}{cccccr@{.}lr@{.}lr@{.}l}
\hline \hline

$C$ & $N_{u,\chi}$ & $N_f^{\rm eN,ee}$ & SMJ & $N_T$ &
\multicolumn{2}{c}{$E_{\rm VMC}$ (a.u.)} & \multicolumn{2}{c}{$\eta$}
& \multicolumn{2}{c}{$\sigma_E^2$ (a.u.)} \\

\hline

2 & 4 & 0 & -- & 10 & $-93$&$90387(3)$  & $93$&$57(7)$\% &
$0$&$645(4)$ \\

2 & 8 & 0 & -- & 18 & $-93$&$90390(5)$  & $93$&$6(1)$\%  &
$0$&$645(4)$ \\

3 & 8 & 0 & -- & 18 & $-93$&$90390(5)$  & $93$&$6(1)$\%  &
$0$&$645(1)$ \\

2 & 4 & 3 & T  & 29 & $-93$&$90672(1)$  & $99$&$81(2)$\% &
$0$&$0810(3)$ \\

2 & 4 & 3 & F  & 37 & $-93$&$90672(2)$  & $99$&$81(4)$\% &
$0$&$0138(8)$ \\

3 & 6 & 3 & F  & 41 & $-93$&$906801(6)$ & $99$&$99(1)$\% &
$0$&$00276(7)$ \\

\hline \hline
\end{tabular}
\caption{Optimized Jastrow factors and VMC energies for Ne$^{8+}$.
The HF energy is $-93.86111347$\,a.u., the exact energy is
$-93.906806$\,a.u., and the DMC energy is within error bars of the
exact value.  Duplication of $u$ and $\chi$ by $f$ is permitted.  In
each case $S_u=S_f=S_\chi=0$, $N_u=N_\chi \equiv N_{u,\chi}$, and
$N_f^{\rm eN} = N_f^{\rm ee} \equiv N_f^{\rm eN,ee}$.
\label{table:varmin_results_Ne8+}}
\end{center}
\end{table}

Tables~\ref{table:varmin_results_He}
and~\ref{table:varmin_results_Ne8+} show variational energies of
optimized Jastrow factors for He and Ne$^{8+}$.  When using $u$ and
$\chi$ functions only we obtain $91.17(2)$\% (He) and $93.64(11)$\%
(Ne$^{8+}$) of the correlation energy, but when we add an $f$ term we
obtain nearly 100\% of the correlation energy.  Elimination of the
terms of the forms $r_i h(r_j,r_{ij})$ and $r_{ij} g(r_i,r_j)$ leads
to an expression containing the same powers as the Jastrow factor of
Schmidt and Moskowitz.\cite{schmidt_1990,moskowitz_1992} The
additional terms are unimportant in He and Ne$^{8+}$.  The results are
not strongly dependent on whether $C=2$ or 3, or whether duplication
of the terms in $u$ and $\chi$ by those in $f$ is prevented or not.

Our results for Ne$^{8+}$ are better than our results for He, both
with and without the $(r_i,r_j,r_{ij})$ terms in the Jastrow factor.
This is to be expected, because the electron--electron interaction is
a smaller perturbation in Ne$^{8+}$, and hence correlation effects are
less significant.

Using a Jastrow factor consisting of a fourth-order Pad\'e function of
scaled variables, Umrigar \textit{et al.}~\cite{umrigar_1988a}
obtained a VMC energy of $-2.903726(4)$\,a.u.\ for He, so they were
able to retrieve 100\% of the correlation energy in this case.  Making
use of scaled variables, instead of cutting off the Jastrow factor at
a finite range, would therefore appear to be beneficial in the special
case of two-electron atoms.

In general we find that the Jastrow factors which recover a large
fraction of the correlation energy have a correspondingly low
variance.  However, the variance obtained for Ne$^{8+}$ using
``SMJ''~=~True is surprisingly high, even though the variational
energy is about the same as the corresponding result in which the full
variational freedom of $f$ is used.

\subsection{All-electron Ne and pseudo-Ne}

\begin{table}
\begin{center}
\begin{tabular}{cccccccr@{.}lr@{.}lr@{.}l}
\hline \hline

$C$ & $N_\chi$ & $N_f^{\rm eN,ee}$ & $S_f$ & D & SMJ & $N_T$ &
\multicolumn{2}{c}{$E_{\rm VMC}$ (a.u.)} & \multicolumn{2}{c}{$\eta$}
& \multicolumn{2}{c}{$\sigma_E^2$ (a.u.)} \\

\hline

2 & 8 & 0 & -- & -- & -- & 26 & $-128$&$757(9)$    &
\multicolumn{2}{l}{$56(2)$\%} & \,$3$&$17(6)$ \\

2 & 0 & 2 & 0  & F  & F  & 23 & $-128$&$781(9)$    &
\multicolumn{2}{l}{$62(2)$\%} & $3$&$2(1)$ \\

3 & 0 & 2 & 0  & T  & F  & 26 & $-128$&$850(7)$    &
\multicolumn{2}{l}{$80(2)$\%} & $2$&$14(6)$ \\

2 & 0 & 2 & 0  & T  & F  & 26 & $-128$&$863(7)$    &
\multicolumn{2}{l}{$84(2)$\%} & $2$&$2(1)$ \\

2 & 0 & 3 & 0  & F  & F  & 39 & $-128$&$868(7)$    &
\multicolumn{2}{l}{$85(2)$\%} & $2$&$03(1)$ \\

2 & 4 & 3 & 0  & T  & T  & 41  & $-128$&$876(2)$   & $87$&$3(6)$\% &
$1$&$92(1)$ \\

2 & 0 & 3 & 0  & T  & F  & 44 & $-128$&$877(6)$    &
\multicolumn{2}{l}{$88(2)$\%} & $1$&$49(3)$ \\

2 & 4 & 3 & 0  & T  & F  & 49 & $-128$&$886(2)$  & $90$&$0(6)$\% &
$1$&$27(2)$ \\

2 & 4 & 3 & 1  & T  & F  & 75 & $-128$&$8983(2)$ & $93$&$2(2)$\% &
$1$&$12(2)$ \\

\hline \hline
\end{tabular}
\caption{Optimized Jastrow factors and VMC energies for all-electron
Ne.  The HF energy is $-128.54709807$\,a.u., the exact energy is
$-128.9376$\,a.u.,\cite{davidson_1991,chakravorty_1993} and our DMC
energy is $-128.9238(7)$\,a.u.  In all cases $S_u=1$, $S_\chi=0$,
$N_u=8$, and $N_f^{\rm eN}=N_f^{\rm eN} \equiv N_f^{\rm eN,ee}$.
\label{table:varmin_results_newjas_Ne_AE}}
\end{center}
\end{table}

The results of optimizing different Jastrow factors for the
all-electron Ne atom are given in Table
\ref{table:varmin_results_newjas_Ne_AE}.  The importance of the $f$
terms is clear: less than 60\% of the correlation energy can be
retrieved using only $u$ and $\chi$, whereas more than 90\% can be
retrieved if $f$ terms are used as well.

We find that using $C=2$ gives slightly better results than $C=3$: it
does not cause our optimization procedure any difficulties, and the
extra variational freedom can be exploited in this case.  The
discontinuities in the local energy do not appear to cause any
population-control problems for the DMC algorithm either.

The optimal values of the cutoff lengths $L_u$, $L_\chi$, and $L_f$
lie between 2\,a.u.\ and 3\,a.u.\ in most cases.  In our best wave
functions $L_u$ is the longest of the three.  Where $\chi$ is absent,
however, $f$ has the greatest cutoff length.  We tried optimizing more
than one $f$ function in order to allow separate $(r_i,r_j,r_{ij})$
correlations for the core and valence electrons, but this did not
lower the variational energy.

If $N_f^{\rm eN}=N_f^{\rm ee}=2$ and $\chi$ is absent then it is
important to allow $f$ to duplicate $u$ and $\chi$.  62(2)\% of the
correlation energy is retrieved when duplication is disallowed whereas
84(2)\% is retrieved when duplication is permitted.  However, the
difference is far less pronounced when $N_f^{\rm eN}=N_f^{\rm ee}=3$:
about 85\% is retrieved irrespective of whether duplication is
allowed.  Using $N_f^{\rm eN}=N_f^{\rm ee}=2$ and allowing duplication
of $u$ and $\chi$ gives a more efficient parameterization of the
Jastrow factor, for the number of parameters is substantially less
than is the case when $N_f^{\rm eN}=N_f^{\rm ee}=3$.  In these
calculations the optimal cutoff length of $f$ ($L_f \simeq
3.7$\,a.u.)\ is greater than that of $u$ ($L_u \simeq 1.0$\,a.u.).
Isolated atoms are a special case in which the $\chi$ function can be
long-ranged.  In the absence of $\chi$, $f$ is forced to be
long-ranged so that it can describe the electron--nucleus
correlations.  Hence $u$, rather than $f$, has to describe all the
short-ranged electron--electron correlations.

We obtain slightly better results when we include the terms in $f$
that are neglected in the Schmidt--Moskowitz Jastrow factor.  The VMC
energy is fairly insensitive to the spin-dependence of $f$.

We have investigated whether it is better to include the
electron--nucleus cusp in the Jastrow factor or in the orbitals in the
Slater wave function.  Calculations were carried out using orbitals
expanded in a Gaussian basis set, generated by the \textsc{crystal}
code.\cite{crystal_98} The $\chi$ term in the Jastrow factor satisfied
the electron--nucleus cusp condition.  The results obtained were
significantly poorer than those shown in Table
\ref{table:varmin_results_newjas_Ne_AE}.  In order to get reasonable
variational energies, a very large number of $\chi$ parameters was
required, with $N_\chi \geq 15$.  Even with $N_\chi=15$, only about
25\% of the correlation energy was retrieved.  It is clearly
preferable to use orbitals that satisfy the electron--nucleus cusp
condition.

There is a significant fixed-node error in the DMC energy: our DMC
energy is $0.0138(7)$\,a.u.\ higher than the exact non-relativistic
ground-state energy.\cite{davidson_1991,chakravorty_1993} We have
verified that population-control biases are negligibly small and we
have performed an extrapolation to zero time step, so the only
remaining bias in our DMC energy is the fixed-node error.  The best
all-electron VMC energy reported in the literature is that of Huang
\textit{et al.},\cite{huang_1997} who optimized parameters in their
orbitals at the same time as their Jastrow factor, giving them extra
variational freedom, including the opportunity to reduce the
fixed-node error. Using a Jastrow factor containing the same types of
correlation as ours (electron--electron, electron--nucleus, and
electron--electron--nucleus), and optimizing the orbitals as well as
the Jastrow factor, Huang \textit{et al.}\ obtain a VMC energy of
$-128.9008(1)$\,a.u., which is only slightly lower than our best
energy of $-128.8983(2)$\,a.u.

\begin{table}
\begin{center}
\begin{tabular}{cccr@{.}lr@{.}lr@{.}l}
\hline \hline

$N_f^{\rm eN,ee}$ & D & $N_T$ & \multicolumn{2}{c}{$E_{\rm VMC}$
(a.u.)} & \multicolumn{2}{c}{$\eta$} & \multicolumn{2}{c}{$\sigma_E^2$
(a.u.)} \\

\hline

0 & -- & 26 & $-34$&$879(1)$ & $86$&$2(3)$\% & \,$0$&$74(2)$ \\

3 & F  & 48 & $-34$&$904(1)$ & $94$&$2(3)$\% & $0$&$45(1)$ \\

2 & F  & 32 & $-34$&$905(1)$ & $94$&$5(3)$\% & $0$&$51(3)$ \\

2 & T  & 35 & $-34$&$908(1)$ & $95$&$5(3)$\% & $0$&$460(4)$ \\

\hline \hline
\end{tabular}
\caption{Optimized Jastrow factors and VMC energies for pseudo-Ne.  An
  HF pseudopotential was used to represent the Ne$^{8+}$
  ion.\cite{trail_2004} The HF energy is $-34.6105$\,a.u.\ and the DMC
  energy is $-34.9220(4)$\,a.u.  In all cases $C=2$, $S_u=1$,
  $S_\chi=S_f=0$, $N_u=N_\chi=8$, and $N_f^{\rm eN}=N_f^{\rm ee}
  \equiv N_f^{\rm eN,ee}$.
\label{table:varmin_results_newjas_Ne_PsP}}
\end{center}
\end{table}

The results of optimizing Jastrow factors for pseudo-Ne are shown in
Table~\ref{table:varmin_results_newjas_Ne_PsP}.  $f$ is much less
important in the pseudo-atom than in all-electron Ne.  $86.3(5)$\% of
the correlation energy is retrieved using $u$ and $\chi$ only, while
$95.7(4)$\% is retrieved when $f$ is used too.  A greater fraction of
the correlation energy can be retrieved in the pseudo-atom than in the
all-electron atom.

\section{Example II: SiH$_4$ molecule}
\label{section:silane}

We used a bond length of $2.8289$\,a.u.\ for the SiH$_4$ (silane)
molecule, in which the Si$^{4+}$ ion was represented by a relativistic
HF pseudopotential~\cite{lee_thesis}   and the full Coulomb potential
was used for the hydrogen nuclei.  The orbitals forming the Slater
determinants were obtained from HF calculations using a large Gaussian
basis set and the \textsc{gaussian} code.\cite{gaussian_98}

\begin{table}
\begin{center}
\begin{tabular}{ccccr@{.}lcr@{.}l}
\hline \hline

$C$ & $N_{u,\chi}$ & $N_f^{\rm eN,ee}$ & $N_T$ &
\multicolumn{2}{c}{$E_{\rm VMC}$ (a.u.)} & $\eta$ &
\multicolumn{2}{c}{$\sigma_E^2$ (a.u.)} \\

\hline

2 &  1 & 0 & 7  & $-6$&$284(2)$  & $88(1)$\% & \,$0$&$096(2)$ \\

2 & 12 & 0 & 51 & $-6$&$291(2)$  & $92(1)$\% & $0$&$066(4)$ \\

3 &  4 & 0 & 19 & $-6$&$291(2)$  & $92(1)$\% & $0$&$07(1)$ \\

3 &  4 & 2 & 31 & $-6$&$292(2)$  & $92(1)$\% & $0$&$08(2)$ \\

2 &  4 & 0 & 19 & $-6$&$293(2)$  & $93(1)$\% & $0$&$067(5)$ \\

\hline \hline
\end{tabular}
\caption{Optimized Jastrow factors and VMC energies for SiH$_4$.  The
HF energy is $-6.118$\,a.u.\ and the DMC energy is $-6.3064(2)$\,a.u.
In each case $S_u=1$, $S_\chi=S_f=0$, $N_u=N_\chi \equiv N_{u,\chi}$,
and $N_f^{\rm eN}=N_f^{\rm ee} \equiv N_f^{\rm eN,ee}$.  Duplication
of $u$ and $\chi$ by $f$ is prohibited.
\label{table:varmin_results_silane}}
\end{center}
\end{table}

Results for some of the Jastrow factors tested for SiH$_4$ are given
in Table~\ref{table:varmin_results_silane}. We find that a large
fraction of the correlation energy can be obtained using rather simple
Jastrow factors.  Using $u$ and $\chi$ functions only, and with a
total of only 7 parameters, we are able to obtain almost 90\% of the
correlation energy.  Our best Jastrow factors obtain about 93(1)\% of
the correlation energy.  We find the optimal cutoff lengths ($L_{u}
\simeq 10$ a.u.\ and $L_{\chi {\rm Si}} \simeq L_{\chi {\rm H}} \simeq
5$ a.u.)\ to be fairly independent of $N_u$ and $N_\chi$.  There is no
detectable benefit from going beyond $N_u=N_\chi=4$, or from
introducing $f$ functions.  Both the results obtained and the behavior
of the optimization procedure are very similar for $C=2$ and $C=3$, so
that in this case there is no benefit from having a continuous local
energy.

\section{Example III: Ni atom and NiO dimer}
\label{section:Ni}

We investigated the Ni atom and the NiO dimer with a bond length of
$3.075$\,a.u., using HF pseudopotentials~\cite{trail_2004} to
represent the Ni$^{10+}$ and O$^{6+}$ ions.  The orbitals were
obtained from HF calculations using a large Gaussian basis set and the
\textsc{crystal} code.\cite{crystal_98} We find that the $f$ functions
are significant for both the Ni atom
(Table~\ref{table:varmin_results_Ni}) and the NiO dimer
(Table~\ref{table:varmin_results_NiO}) in spite of the use of
pseudopotentials.

\begin{table}
\begin{center}
\begin{tabular}{ccccccr@{.}lr@{.}lr@{.}l}
\hline \hline

$N_f^{\rm eN,ee}$ & $S_u$ & $S_\chi$ & $S_f$ & D & $N_T$ &
    \multicolumn{2}{c}{$E_{\rm VMC}$ (a.u.)} &
    \multicolumn{2}{c}{$\eta$} & \multicolumn{2}{c}{$\sigma_E^2$
    (a.u.)} \\

\hline

0 & 1 & 0 & -- & -- & 26  & $-39$&$0598(4)$ & $69$&$65(9)$\% &
$1$&$148(2)$ \\

0 & 1 & 1 & -- & -- & 34  & $-39$&$0673(4)$ & $70$&$98(9)$\% &
$1$&$114(3)$ \\

0 & 2 & 1 & -- & -- & 42  & $-39$&$0719(4)$ & $71$&$8(1)$\%  &
$1$&$098(4)$ \\

3 & 1 & 0 & 0  & F  & 48  & $-39$&$1045(3)$ & $77$&$57(9)$\% &
$0$&$721(3)$ \\

3 & 1 & 1 & 0  & F  & 56  & $-39$&$1074(3)$ & $78$&$09(9)$\% &
$0$&$695(1)$ \\

3 & 2 & 1 & 0  & F  & 64  & $-39$&$1231(3)$ & $80$&$87(9)$\% &
$0$&$734(1)$ \\

3 & 2 & 1 & 1  & F  & 85 & $-39$&$1247(3)$  & $81$&$15(9)$\% &
$0$&$743(2)$ \\

3 & 2 & 1 & 2  & F  & 106 & $-39$&$1247(3)$ & $81$&$15(9)$\% &
$0$&$727(4)$ \\

3 & 2 & 1 & 2  & T  & 121 & $-39$&$1469(2)$ & $85$&$09(8)$\% &
$0$&$571(1)$ \\

\hline \hline
\end{tabular}
\caption{Optimized Jastrow factors and VMC energies for pseudo-Ni.
The HF energy is $-38.6670$\,a.u.\ and the DMC energy is
$-39.2310(5)$\,a.u.  In each case $C=3$, $N_u=N_\chi=8$ and $N_f^{\rm
eN}=N_f^{\rm ee} \equiv N_f^{\rm eN,ee}$.
\label{table:varmin_results_Ni}}
\end{center}
\end{table}

\begin{table}
\begin{center}
\begin{tabular}{ccccr@{.}lr@{.}lr@{.}l}
\hline \hline

$N_f^{\rm eN,ee}$ & $S_u$ & $S_{\chi,f}$ & $N_T$ &
\multicolumn{2}{c}{$E_{\rm VMC}$ (a.u.)} & \multicolumn{2}{c}{$\eta$}
& \multicolumn{2}{c}{$\sigma_E^2$ (a.u.)} \\

\hline

0 & 1 & 0 & 35  & $-54$&$9031(5)$ & $69$&$99(8)$\% & $2$&$144(4)$ \\

0 & 1 & 1 & 51  & $-54$&$9069(5)$ & $70$&$45(8)$\% & $2$&$132(4)$ \\

3 & 1 & 0 & 77  & $-54$&$9984(4)$ & $81$&$31(7)$\% & $1$&$521(3)$ \\

3 & 1 & 1 & 157 & $-55$&$0104(4)$ & $82$&$74(8)$\% & $1$&$451(2)$ \\

3 & 2 & 1 & 165 & $-55$&$0105(4)$ & $82$&$75(8)$\% & $1$&$439(2)$ \\

\hline \hline
\end{tabular}
\caption{Optimized Jastrow factors and VMC energies for the NiO dimer.
The HF energy is $-54.31362$\,a.u.\ and the DMC energy is
$-55.1558(6)$\,a.u.  In each case $C=3$, $S_u=1$, $S_\chi=S_f \equiv
S_{\chi,f}$, $N_u=N_\chi=8$, and $N_f^{\rm eN}=N_f^{\rm ee} \equiv
N_f^{\rm eN,ee}$.  Duplication of $u$ and $\chi$ by $f$ is forbidden.
\label{table:varmin_results_NiO}}
\end{center}
\end{table}

Note that Ni and NiO are partially spin-polarized, so that it may be
advantageous to have different $\chi$ functions for spin-up and
spin-down electrons, and different $u$ and $f$ functions for parallel
spin-up and parallel spin-down pairs of electrons, unlike the other
systems studied in this work.  Our results show that the
spin-dependences of the $u$, $\chi$, and $f$ functions do indeed have
a significant effect on the quality of the wave functions for Ni and
NiO, although including $(r_i,r_j,r_{ij})$ terms in the Jastrow factor
has a greater effect.  An additional 1--2\% of the correlation energy
can be retrieved when $\chi$ and $u$ are allowed to differ for spin-up
and spin-down electrons. Using different $f$ functions for
antiparallel, parallel spin-up, and parallel spin-down pairs also
lowers the variational energy slightly, although it greatly increases
the number of parameters which have to be optimized.

These calculations are the only ones for which we have retrieved less
than 90\% of the correlation energy.

\section{Example IV: Si solid}
\label{section:Si}

\subsection{16-atom simulation cell \label{section:si_2x2x2}}

We modeled crystalline Si in the diamond structure using a 16-atom,
face-centered cubic simulation cell subject to periodic boundary
conditions.  We used a cubic lattice constant of $5.12966$\,a.u., and
the Si$^{4+}$ ions were represented by
pseudopotentials.\cite{lee_thesis}  The orbitals were obtained from HF
calculations using a large Gaussian basis set and the \textsc{crystal}
code.\cite{crystal_98} The results of optimizing the our Jastrow
factor are shown in Table \ref{table:varmin_results_newjas_Si}.

$L_\chi$ adjusts itself to sizes of the order of the inter-atomic
spacing ($L_\chi \simeq 6.4$\,a.u., whereas the nearest-neighbor
distance is $4.4424$\,a.u.), while $L_u$ tends to the largest possible
value, which is the Wigner--Seitz radius of the simulation cell
($7.2544$\,a.u.).

It is much easier to optimize the cutoff lengths when $C=3$ than when
$C=2$.  It seems that the discontinuities in the local energy that are
present when $C=2$ cause serious problems for our optimization
procedures.  For example, a flexible Jastrow factor with $C=2$ gives a
lower variance, but higher energy, than a simple Jastrow factor with
$C=3$.  Such problems were not apparent in our calculations for atoms
and small molecules.  The discontinuities do not appear to lead to any
problems within DMC calculations, however.

The use of the $p$ and $q$ terms does not bring about a statistically
significant lowering of the VMC energy.  The optimal value of $L_u$ is
a little less than the Wigner--Seitz radius when $p$ terms are
included, so the $p$ term must describe the long-ranged correlations,
as expected.  The sinusoidal $p$ and $q$ functions are considerably
more expensive to evaluate than the polynomial $u$ and $\chi$
functions, so it is anticipated that $p$ and $q$ will rarely be used
in practice, except in strongly anisotropic systems such as graphite,
where Prendergast \textit{et al.}~\cite{prendergast} have demonstrated
that plane-wave expansions in the Jastrow factor have an important
role to play.

\begin{table}
\begin{center}
\begin{tabular}{cccccr@{.}lr@{.}lr@{.}l}
\hline \hline

$C$ & $N_{u,\chi}$ & $N_f^{\rm eN,ee}$ & $N_{p,q}$ & $N_T$ &
\multicolumn{2}{c}{$E_{\rm VMC}$ (a.u.)} & \multicolumn{2}{c}{$\eta$}
& \multicolumn{2}{c}{$\sigma_E^2$ (a.u.)} \\

\hline

2 & 1  & 0 & 0 & 5  & $-7$&$8583(5)$ & $82$&$1(2)$\% & \,$1$&$551(3)$
\\

2 & 10 & 0 & 0 & 32 & $-7$&$8714(4)$ & $87$&$0(2)$\% & $0$&$863(2)$ \\

3 & 1  & 0 & 0 & 5  & $-7$&$8761(4)$ & $88$&$8(2)$\% & $1$&$010(6)$ \\

3 & 10 & 0 & 0 & 32 & $-7$&$8809(4)$ & $90$&$6(2)$\% & $0$&$842(6)$ \\

3 & 4  & 0 & 5 & 29 & $-7$&$8816(3)$ & $90$&$8(1)$\% & $0$&$81(1)$  \\

3 & 4  & 2 & 0 & 23 & $-7$&$8832(3)$ & $91$&$4(1)$\% & $0$&$846(2)$ \\

\hline \hline
\end{tabular}
\caption{Optimized Jastrow factors and VMC energies for pseudo-Si
(16-atom simulation cell).  The HF energy is $-7.63946$\,a.u.\ per
primitive cell and the DMC energy is $-7.90600(6)$\,a.u.\ per
primitive cell.  In each case $S_u=S_p=1$, $S_\chi=S_f=S_q=0$,
$N_u=N_\chi \equiv N_{u,\chi}$, $N_f^{\rm eN} = N_f^{\rm ee} \equiv
N_f^{\rm eN,ee}$, and $N_p=N_q \equiv N_{p,q}$.  Duplication of $u$
and $\chi$ by $f$ is allowed.
\label{table:varmin_results_newjas_Si}}
\end{center}
\end{table}

\subsection{54-atom simulation cell}

Similar calculations to those reported in Sec.~\ref{section:si_2x2x2}
were carried out using a 54-atom simulation cell.  The HF energy is
$-7.6792$\,a.u.\ per primitive cell and the DMC energy is
$-7.9555(2)$\,a.u.\ per primitive cell.  The VMC energy obtained using
our Jastrow factor with $N_u=N\chi=4$, $S_u=1$, and $S_\chi=0$ (giving
14 free parameters) is $-7.9331(6)$\,a.u.\ per primitive cell, so
$91.9(2)$\% of the correlation energy is retrieved.  A very similar
fraction of the correlation energy is retrieved in the 16-atom cell
(see Table \ref{table:varmin_results_newjas_Si}).  The VMC energy
variance is $2.92(5)$\,a.u., so the variance per electron is about the
same as for the 16-atom simulation cell.  The optimal value of $L_u$
is again equal to the Wigner--Seitz radius of the simulation cell
($10.882$\,a.u.).  The optimal value of $L_\chi$ remains of order the
atomic size, at $L_\chi \simeq 4.7$\,a.u.  These results indicate that
cutting off the $\chi$ function at sizes of order the inter-atomic
spacing is a valuable improvement to the Jastrow factor.

\section{Conclusions}
\label{section:conclusions}

We have developed and tested a form of Jastrow factor consisting of
electron--electron ($u$), electron--nucleus ($\chi_I$), and
electron--electron--nucleus ($f_I$) terms, and additional
electron-position-dependent ($q$) and
electron--electron-separation-dependent ($p$) terms.  The $u$,
$\chi_I$, and $f_I$ terms are expanded in polynomials and are forced
to go to zero at some cutoff radii.  The $p$ and $q$ terms are
expanded in plane waves.  We have tested our Jastrow factor on atoms,
molecules, and solids, including both all-electron and pseudopotential
atoms.  In most cases our VMC calculations have retrieved over 90\% of
the fixed-node correlation energy.

The variable parameters appear linearly in our Jastrow factor, except
for the cutoff radii of $u$, $\chi_I$, and $f_I$. The linearity in the
variable parameters aids the computational efficiency of the
optimization algorithm.  We have found that it is often beneficial to
make the local energy continuous at the cutoffs when optimizing the
cutoff radii, but that a lower variational energy can be achieved in
some cases when discontinuities are allowed at the cutoffs.

We have investigated the importance of terms in the Jastrow factor
that were neglected by Schmidt and
Moskowitz,\cite{schmidt_1990,moskowitz_1992} but we found them to be
unimportant in the systems studied here.

The electron--electron--nucleus $f_I$ terms were found to be important
in all-electron simulations of the He, Ne$^{8+}$, and Ne atoms.  The
$f_I$ terms are generally less important for pseudo-atoms than
all-electron atoms.  For example, they account for nearly 40\% of the
correlation energy in all-electron Ne, but only about 10\% in
pseudo-Ne.  We found the $f_I$ terms to be significant for pseudo-Ni
and the NiO dimer, but they had little effect in SiH$_4$ or
crystalline Si.  The plane-wave terms $p$ and $q$ were found to be
unimportant in crystalline Si.

We have found that it is preferable to use orbitals which satisfy the
electron--nucleus cusp condition and to require the Jastrow factor to
be cuspless at nuclei rather than to enforce the electron--nucleus
cusp condition through the Jastrow factor.

Overall we found that optimizing the cutoff radii is very important.
In crystalline Si the cutoff for $u$ adjusted itself to the largest
possible value, which is the Wigner--Seitz radius of the simulation
cell.  This indicates that there are significant correlations in Si
extending over many atoms.  We have argued that the long-ranged parts
of the $\chi_I$ and $f_I$ terms do not give useful variational
freedom, and in support of this we found that the corresponding
cutoffs adjusted themselves to sizes of the order of the inter-atomic
spacing.

Although we have optimized our Jastrow factors by minimizing the
variance of the energy, we have measured the accuracy of the Jastrow
factors in terms of the variational energy itself.  If we were to
minimize the variational energy directly then we might get even better
results.  Recently there has been much interest in developing
energy-minimization methods for optimizing trial wave
functions,\cite{filippi_2000,prendergast,li_1999} and we intend to
pursue this avenue further.

\section{Acknowledgments}

We thank Cyrus Umrigar for helpful discussions and John Trail for
providing the all-electron HF orbitals and the pseudopotentials used
in this work. Financial support was provided by the Engineering and
Physical Sciences Research Council (EPSRC), UK\@.  MDT thanks the
Royal Society for a Research Fellowship.  Computational facilities
were provided by the High Performance Computing Facility at the
University of Cambridge.

\appendix

\section{Imposing the constraints on $\gamma$}
\label{appendix:constraints}

Consider a particular set of ions $I$ and a particular spin-pair type.
To impose the symmetry of $f$ under interchange of electrons we work
with $\gamma_{lmnI}$, where $l \geq m$, and then complete the $\gamma$
array by setting $\gamma_{lmnI}=\gamma_{mlnI}$ for $l<m$.  Let ${\bf
x}$ be a vector whose components are each of the $\gamma$ coefficients
with $l \geq m$.  The remaining constraints (no-cusp conditions and,
optionally, no-duplication-of-$u$-and-$\chi$ conditions) may then be
written in matrix form as $A {\bf x}=0$.  The total number of
constraints determines the number of rows of $A$:
\begin{itemize}
\item There are $2N_{f I}^{\rm eN}+1$ constraints (one for each value
  of $k=l+m$) associated with the imposition of the electron--electron
  no-cusp condition.  Using the symmetry of $\gamma$,
  Eq.~(\ref{equation:gamma_cond_ee}) can be rewritten so that only
  elements of ${\bf x}$ are involved: $\forall \, k \in \{ 0, \ldots,
  2N_{f I}^{\rm eN} \}$,
\begin{equation}
\sum_{l,m {\rm ~:~} {l+m=k \atop l>m }} 2 \gamma_{lm1I} + \sum_{l {\rm
~:~} 2l=k} \gamma_{ll1I} = 0.
\end{equation}
For each $k$, this equation defines a row of $A$.
\item There are $N_{f I}^{\rm eN}+N_{f I}^{\rm ee}+1$ constraints (one
  for each value of $k^\prime=l+n$) associated with the imposition of
  the electron--nucleus no-cusp condition.
  Eq.~(\ref{equation:gamma_cond_eN}) can be rewritten so that only
  elements of ${\bf x}$ are involved: $\forall \, k^\prime \in \{ 0,
  \ldots, N_{f I}^{\rm eN}+N_{f I}^{\rm ee} \}$,
\begin{equation}
C \gamma_{00k^\prime I} - L_{f I} \gamma_{10k^\prime I} + \sum_{l,n
  {\rm ~:~} {l+n=k^\prime \atop l \geq 1}} \left( C \gamma_{l0nI}-L_{f
  I} \gamma_{l1nI} \right) = 0.
\end{equation}
For each $k^\prime$, this equation defines a row of $A$.
\item If desired, there are $N_{f I}^{\rm ee}$ constraints imposed to
  prevent duplication of $u$.  ($\gamma_{00nI}=0$ $\forall n$.)
\item If desired, there are $N_{f I}^{\rm eN}$ constraints imposed to
  prevent duplication of $\chi$.  ($\gamma_{l00I}=0$ $\forall l$.)
\end{itemize}

Imposing the constraints reduces the number of independent variable
parameters by ${\rm rank}(A)$.  The matrix $A$ is transformed into its
row-reduced echelon form by Gaussian elimination, which allows us to
identify a suitable set of independent variable parameters.

\end{document}